\documentclass[12pt,a4paper]{article}
\usepackage{graphicx}
\usepackage{times}
\def\aap{{A\&A}}

\def\apj{{ApJ}}

\def\mnras{{MNRAS}}

\def\aapr{A\&A~Rev.}

\def \ha {H{$_\alpha$}}
\def \pv {P{\sc v}}

\def \mdotun {$10^{-6}\,{\rm M_{\odot}/yr}$}   
\def \mdot {$\dot M$}

\def \fcl   {$f_{\rm cl}$}

\def \fvel  {$f_{\rm vel}$}

\def \lcep  {$\lambda$~Cep}

\title{\bf A proper description of clumping in hot star winds: the key to obtaining reliable mass-loss rates?}
\author{Jon O. Sundqvist$^1$, Joachim Puls$^1$, Achim Feldmeier$^2$, and Stanley P. Owocki$^3$\\
\vspace{1cm}\\
\normalsize $^1$ Universit\"atssternwarte M\"unchen, Scheinerstr. 1, 81679 M\"unchen, Germany\\ 
\normalsize $^2$ Institut f\"ur Physik und Astronomie,
              Karl-Liebknecht-Strasse 24/25, 14476 Potsdam-Golm, Germany\\
\normalsize $^3$  University of Delaware, Bartol Research Institute, Newark, DE 19716, USA}
\date{\mbox{}}
\begin{document}
\maketitle
\pagestyle{empty}
%
% WE REDEFINE THE plain LaTeX PAGESTYLE !!! 
% THIS PAGESTYLE WILL BE USED FOR THE FIRST PAGE ONLY !
%
\def\bull{\vrule height .9ex width .8ex depth -.1ex}
\makeatletter
\def\ps@plain{\let\@mkboth\gobbletwo
\def\@oddhead{}\def\@oddfoot{\hfil\tiny\bull\quad
``The multi-wavelength view of hot, massive stars''; 39$^{\rm th}$ Li\`ege Int.\ Astroph.\ Coll., 12-16 July 2010 \quad\bull}%
\def\@evenhead{}\let\@evenfoot\@oddfoot}
\makeatother
%
% AND DEFINE OUR MACROS FOR THE REFERENCE LIST
% I.E \beginrefer \refer and \endrefer
%
\def\beginrefer{\section*{References}%
\begin{quotation}\mbox{}\par}
\def\refer#1\par{{\setlength{\parindent}{-\leftmargin}\indent#1\par}}
\def\endrefer{\end{quotation}}
%
% BEGIN THE ABSTRACT CHAPTER WITH \noindent\small, ENCLOSE IT IN A GROUP
% AND BOLDFACE THE TITLE.
% 
{\noindent\small{\bf Abstract:} Small-scale inhomogeneities, or
`clumping', in the winds of hot, massive stars are conventionally
included in spectral analyses by assuming optically thin clumps.  To
reconcile investigations of different diagnostics using this
microclumping technique, very low mass-loss rates must be invoked for
O stars.

Recently it has been suggested that by using the microclumping
approximation one may actually drastically underestimate the mass-loss
rates. Here we demonstrate this, present a new, improved description
of clumpy winds, and show how corresponding models, in a combined UV
and optical analysis, can alleviate discrepancies between previously
derived rates and those predicted by the line-driven wind theory.
Furthermore, we show that the structures obtained in time-dependent,
radiation-hydrodynamic simulations of the intrinsic line-driven
instability of such winds, which are the basis to our current
understanding of clumping, in their present-day form seem unable to
provide a fully self-consistent, simultaneous fit to both UV and
optical lines. The reasons for this are discussed.}

% NOW COMES THE MAIN BODY OF THE ARTICLE
%
\section{Introduction}

The winds from hot, massive stars are described by the radiative
line-driven wind theory (Castor\ et al. 1975), in which the standard
model assumes the wind to be stationary, spherically symmetric, and
homogeneous. Despite its apparent success (e.g., Vink et al.\ 2000),
this model is probably oversimplified. In particular, much evidence
for a time-dependent, small-scale inhomogeneous wind (that is, a
`clumped' wind) has over the past years accumulated, from the
theoretical as well as the observational side (for an overview, see
Hamann et al.\ 2008). In the following, we investigate and discuss the
\textit{indirect} evidence for wind clumping that has arisen from
quantitative spectroscopy aiming to infer mass-loss rates from
observations.

\section{Deriving empirical mass-loss rates from hot star winds}

The main mass-loss diagnostics of OB-star winds are UV resonance
lines, \ha~line emission (and other recombination lines), and
infra-red and radio continuum emission. Recently, X-ray emission lines
have also been added to the set (e.g., Cohen et al.\ 2010).
%(e.g., Oskinova et al.\ 2006; Cohen et al.\ 2010). 
Here we will focus on resonance lines and \ha, discussing the
influence of optically thick clumping on these diagnostics and using
the well-studied Galactic O supergiant \lcep~as a test bed.

\subsection{Smooth wind models and microclumping}

When smooth wind models are used, mass-loss rates inferred for a given
star, but from different diagnostics, can vary substantially. For
example, in an analysis by means of the unified NLTE model atmosphere
code {\sc fastwind} (Puls et al.\ 2005), we find that the phosphorus
{\sc v} (\pv) UV resonance lines suggest a mass-loss rate
approximately 20 times lower than the one required for a decent fit of
the \ha~emission. That is, depending on which diagnostic is used, the
inferred mass-loss rate of \lcep~can vary by more than an order of
magnitude. This inconsistency has been interpreted as a consequence of
neglecting clumping when deriving these rates.

Wind clumping has traditionally been included in diagnostic tools by
assuming statistically distributed \textit{optically thin} clumps and
a void inter-clump medium, while keeping a smooth velocity field.  The
main result of this \textit{microclumping} approach is that mass-loss
rates derived from smooth models and diagnostics that depend on the
square of the density (such as \ha~in OB-star winds) must be scaled
down by the square root of the clumping factor, \fcl(r)\,$\equiv
\langle \rho^2 \rangle/ \langle \rho \rangle^2$ with the angle
brackets denoting spatially averaged quantities. On the other hand,
processes that depend linearly on the density (such as the UV
resonance lines) are not directly affected by microclumping.  Thus,
with such an optically thin clump model, the only way to reconcile the
above large deviations in rates would be to accept the very low
mass-loss rate indicated by PV, and so assume very high clumping
factors (in this case \fcl\,$\approx$\,400!) to simultaneously
reproduce \ha. Using this technique, we would derive a mass-loss rate
for \lcep~that is roughly an order of magnitude lower than predicted
by the line-driven wind theory (according to the mass-loss recipe in
Vink et al.\ 2000). Similar results have been found by, e.g., Bouret
et al.\ (2005) and Fullerton et al.\ (2006).

Such low mass-loss rates would have enormous implications for massive
star evolution and feedback, and of course also cast severe doubts on
the validity of the line-driven wind theory. In addition, the extreme
clumping factors we had to invoke to reproduce \ha~are in stark
contrast with theoretical predictions (e.g., Runacres \& Owocki 2002,
see also Fig.~\ref{Fig:profs}).  A possible solution of this dilemma
is that clumps are not optically thin for the diagnostic lines under
consideration, which would lead to underestimated rates if
microclumping were still assumed (Oskinova et al.\ 2007; Owocki 2008;
Sundqvist et al.\ 2010a).

\subsection{Relaxing the microclumping approximation} 
\label{tau_cl}

\begin{figure}[h]
\centering
\includegraphics[angle=90,width=5.0cm]{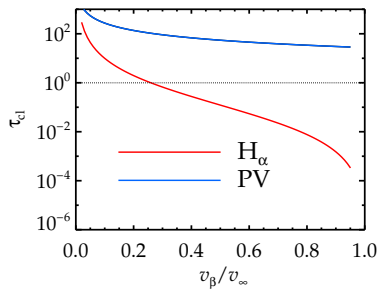}
\caption{Clump optical depths for \ha~and \pv~line formation as functions 
of wind velocity. Stellar and wind parameters as for \lcep, see text.}
\label{Fig:tau_cl}
\end{figure}

\paragraph{Clump optical depths.}
The critical parameter determining the validity of the microclumping
approximation is the \textit{clump optical depth} (which of course is
different for different diagnostics). Actually, for both resonance
lines and \ha, we may quite easily obtain reasonable estimates for
this quantity. In the following, all radii are given in units of the
stellar radius and all velocities in units of the terminal speed. The
\textit{radial} clump optical depth for a resonance line may
then, in the Sobolev approximation, be written as
\begin{equation} 
  \tau_{\rm cl}^{\rm res} \approx \frac{\tau_{\rm sm}^{\rm res}}{f_{\rm vel}} \approx  \frac{\kappa_{\rm 0}q f_{\rm cl}}{r^2 v {\rm d}v/{\rm d}r}, 
  \label{Eq:dtau}
\end{equation}
where $q$ is the ionization fraction of the considered ion and $f_{\rm
  vel}$ a \textit{velocity} filling factor (Owocki 2008), defined as
the ratio between the clump velocity span, $\delta v$, to their
velocity separation, in full analogy with the traditional volume
filling factor. $f_{\rm vel}$ largely controls how a perturbed
velocity field affects the line formation, and may be approximated by
$f_{\rm vel} \approx |\delta v/\delta v_\beta|f_{\rm cl}^{-1}$
(Sundqvist et al.\ 2010b), where $\delta v_\beta$ is the clump
velocity span of a corresponding model with a smooth velocity field. 
As seen from the middle term in Eq.~\ref{Eq:dtau}, \fvel~also 
controls how the clump optical
depth differs from the corresponding smooth one.  $\kappa_0$ is a
line-strength parameter proportional to the mass-loss rate and the
abundance of the considered element, and taken to be constant within
the wind (e.g., Puls et al.\ 2008). As an example, assuming a
mass-loss rate 3.0\,$\times$\,\mdotun~and a solar phosphorus
abundance, we get $\kappa_0 \approx 3.0$ for the blue component of
\pv~in \lcep.  Note that Eq.~\ref{Eq:dtau} assumes that the clumps
cover a complete resonance zone. This is reasonable for the wind's
inner parts, but may be questionable for its outer, slowly
accelerating parts (see Sundqvist et al.\ 2010b). For our discussion
here, however, a Sobolev treatment suffices.
For \ha, the analogy to $\kappa_0$ is the parameter $A$ (Puls et
al.\ 1996)\footnote{$A$ may be readily modified to handle other
  recombination lines than \ha.}, which is proportional to the
mass-loss rate \textit{squared} and to the departure coefficient of
the lower transition level. We may write the radial Sobolev clump
optical depth for \ha~as
\begin{equation} 
  \tau_{\rm cl}^{\rm rec} \approx \tau_{\rm sm}^{\rm rec}\frac{f_{\rm cl}}{f_{\rm vel}} \approx A \frac{f_{\rm cl}^2}{r^4 v^2 {\rm d}v/{\rm d}r}.
  \label{Eq:dtau2}
\end{equation}
Assuming unity departure coefficients, and a mass-loss rate as above,
$A \approx 1.5 \times 10^{-3}$ for \ha~in \lcep.  Note the extra
\fcl~term in this expression, stemming from the $\rho^2$-dependence of
this diagnostic. In Fig~\ref{Fig:tau_cl}, we plot the \ha~and
\pv~clump optical depths, using $\kappa_0$ and $A$ as just given and
\fcl\,=\,9, which is a typical value. For simplicity, we also used
$q$\,=\,1 for the \pv~line formation and a smooth velocity field, $v =
1-0.99/r$ (i.e., a `$\beta$\,=1' law), which gives
\fcl\,=\fvel$^{-1}$, when calculating the clump optical depths
displayed in the figure. Clearly, within these
approximations the clumps remain optically thick for \pv~throughout
the entire wind. For \ha, the $\rho^2$-dependence makes the optical
depth decrease faster with increasing radii, so this line is optically
thick only in the lower wind.  Based only on these simple estimates,
we may therefore expect that resonance lines such as \pv~should be
sensitive to deviations from microclumping in the complete line
profile, whereas recombination lines such as \ha~should be affected
only in the line core. We note also that even if we were to reduce the
ion fraction of \pv~to, say, $q=0.1$, the clumps would still be
optically thick. This illustrates the necessity of relaxing the
microclumping approximation for typical mass-loss line diagnostics of
hot star winds.

\paragraph{Wind models.} 
To investigate effects on the line formation from optically thick
clumps, a non-void inter-clump medium, and a non-monotonic velocity
field, we create (pseudo-)2D and 3D wind models, following the `patch
method' of Dessart \& Owocki 2002. In this technique, the full wind is
`patched' together by assembling inhomogeneous, spherically symmetric,
wind-snapshots in radially independent slices. We construct winds
using both self-consistent, time-dependent, radiation-hydrodynamic
(RH) simulations (computed following Feldmeier et al.\ 1997) and
empirical, stochastic models. Line synthesis of resonance lines is
carried out using the Monte-Carlo method described in Sundqvist et
al.\ (2010a), whereas for \ha~we have developed a new radiative
transfer code, in which we solve the `formal integral' within our 3D
winds, assuming that the departure coefficients are unaffected by
optically thick clumping, which should be reasonable for the O-star
winds discussed here (Sundqvist et al.\ 2010b). Hydrogen departure
coefficients, and ionization fractions of \pv, are calculated with
{\sc fastwind}, accounting for radially dependent microclumping.
Stellar rotation is treated by the standard convolution procedure of a
constant $v \sin i$ (thus neglecting differential rotation), set to 
220\,$\rm km\,s^{-1}$.

When creating our stochastic models, we take an heuristic approach and
use a set of parameters to define the structured medium. If clumps are
optically thick for the investigated diagnostic, the line formation
will generally depend on more structure parameters than just
\fcl. These parameters (for a two component medium) were defined and
discussed in Sundqvist et al.\ (2010a). They are the inter-clump
medium density, the physical distances between the clumps, $v_\beta
\delta t$, set by the time interval $\delta t$ between the release of
two clumps and in our geometry equal to the porosity length (Owocki et
al.\ 2004), and finally the velocity filling factor, $f_{\rm vel}$
(see previous paragraph). We stress that these parameters are
essential for the radiative transfer in an inhomogeneous medium with
optically thick clumps, and not merely `ad-hoc parameters' used in a
fitting procedure. We notice also that in our stochastic models they
are used to \textit{define} the structured wind, and so are
independent of the origin to the inhomogeneities.  This should be
distinguished from the RH simulations, in which the structure arises
naturally from following the time evolution of the wind and stems
directly from the line-driven instability. For these simulations then,
the averaged structure parameters are an \textit{outcome}.

\section{Results from an exemplary study of \lcep}

\begin{figure}[h]
\centering
\begin{minipage}{5.5cm}
\resizebox{\hsize}{!}
{\includegraphics[angle=90]{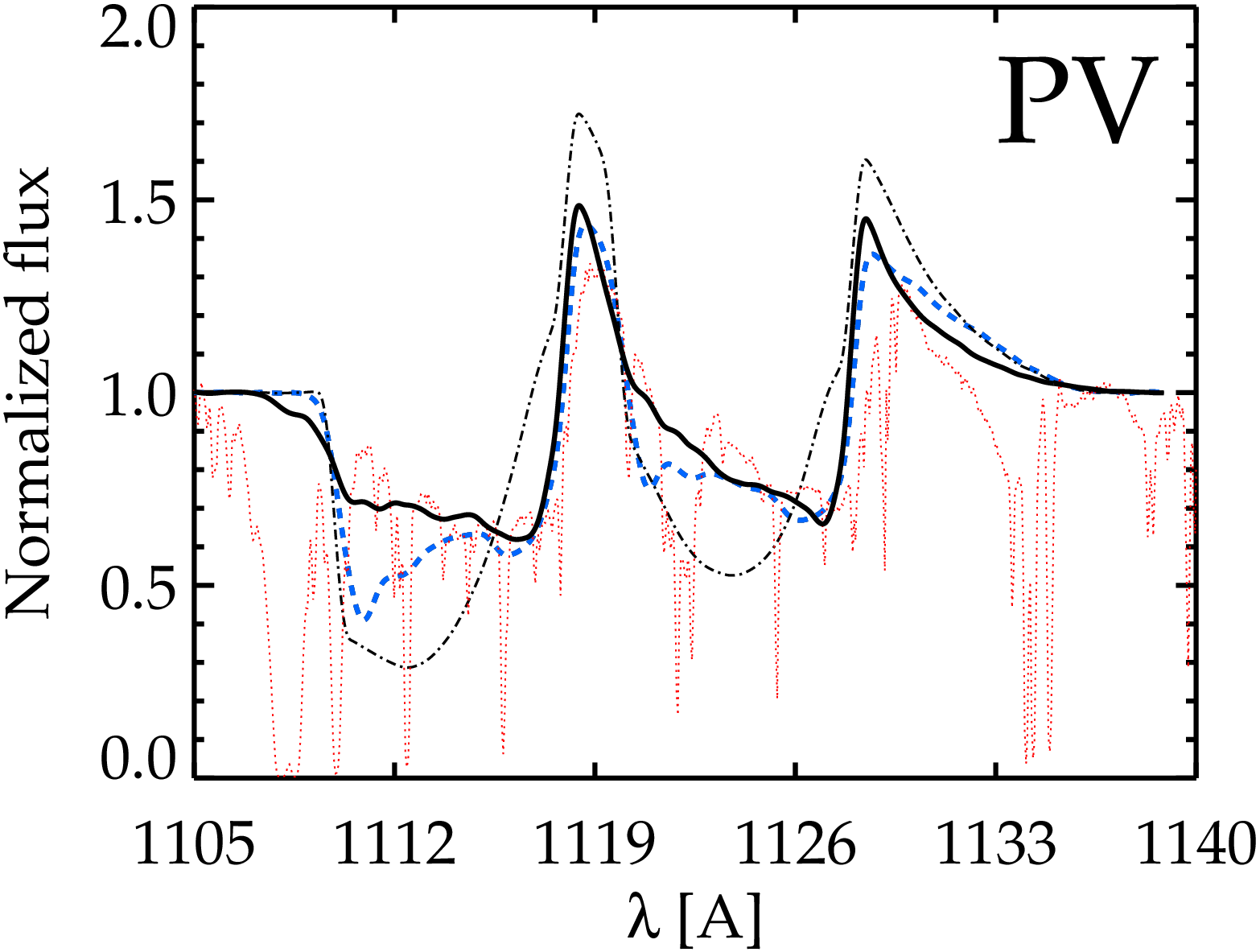}}
%\caption{This is the caption of the left-hand side figure. \label{Fig:}}
\end{minipage}
%\hfill
%\hspace{0.5cm}
\begin{minipage}{5.5cm}
%\centering
\resizebox{\hsize}{!}
{\includegraphics[angle=90]{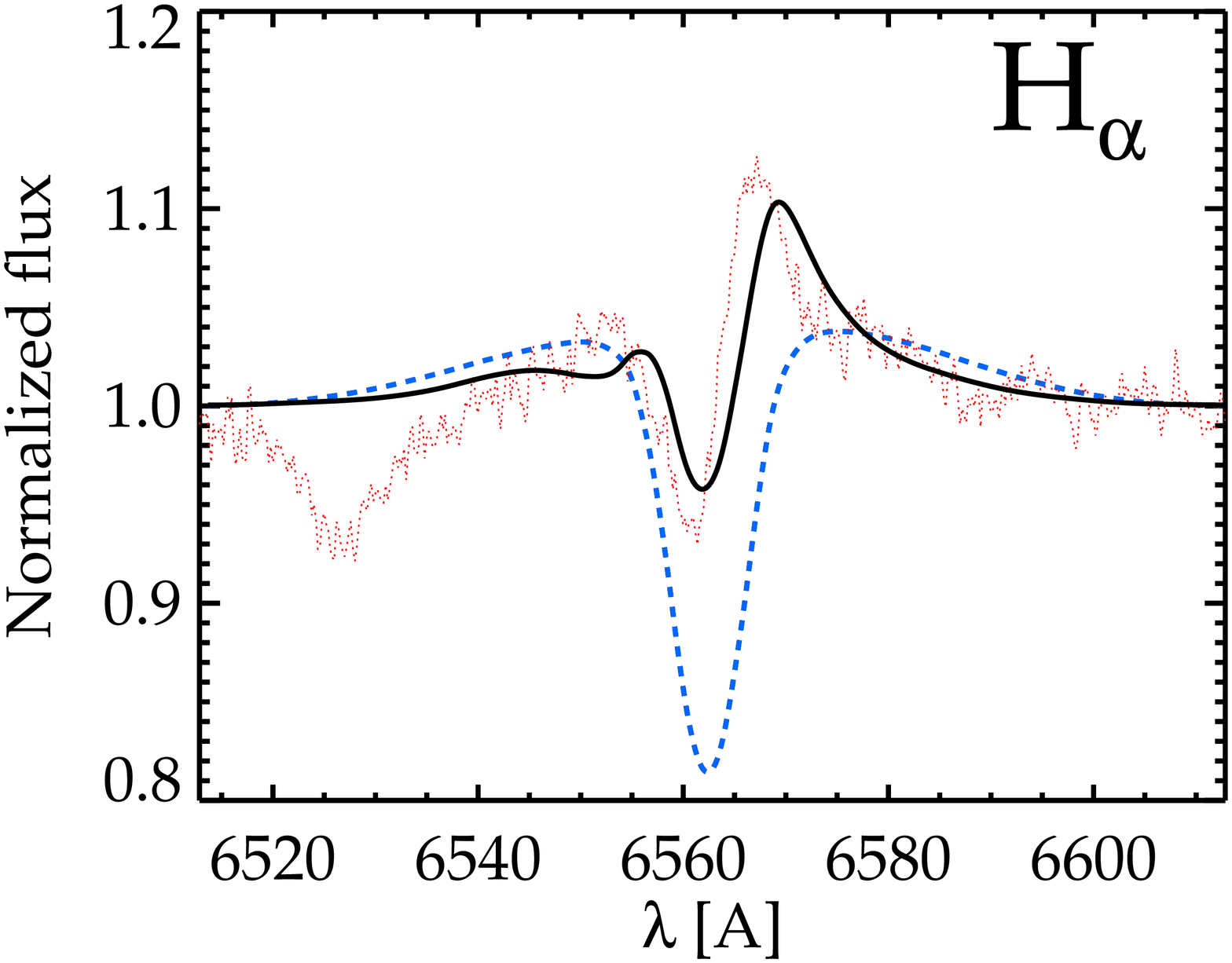}}
%\caption{This is the caption of the right-hand side figure. \label{fig_3}}
\end{minipage}
\begin{minipage}{5.5cm}
%\centering
\resizebox{\hsize}{!}
{\includegraphics[angle=90]{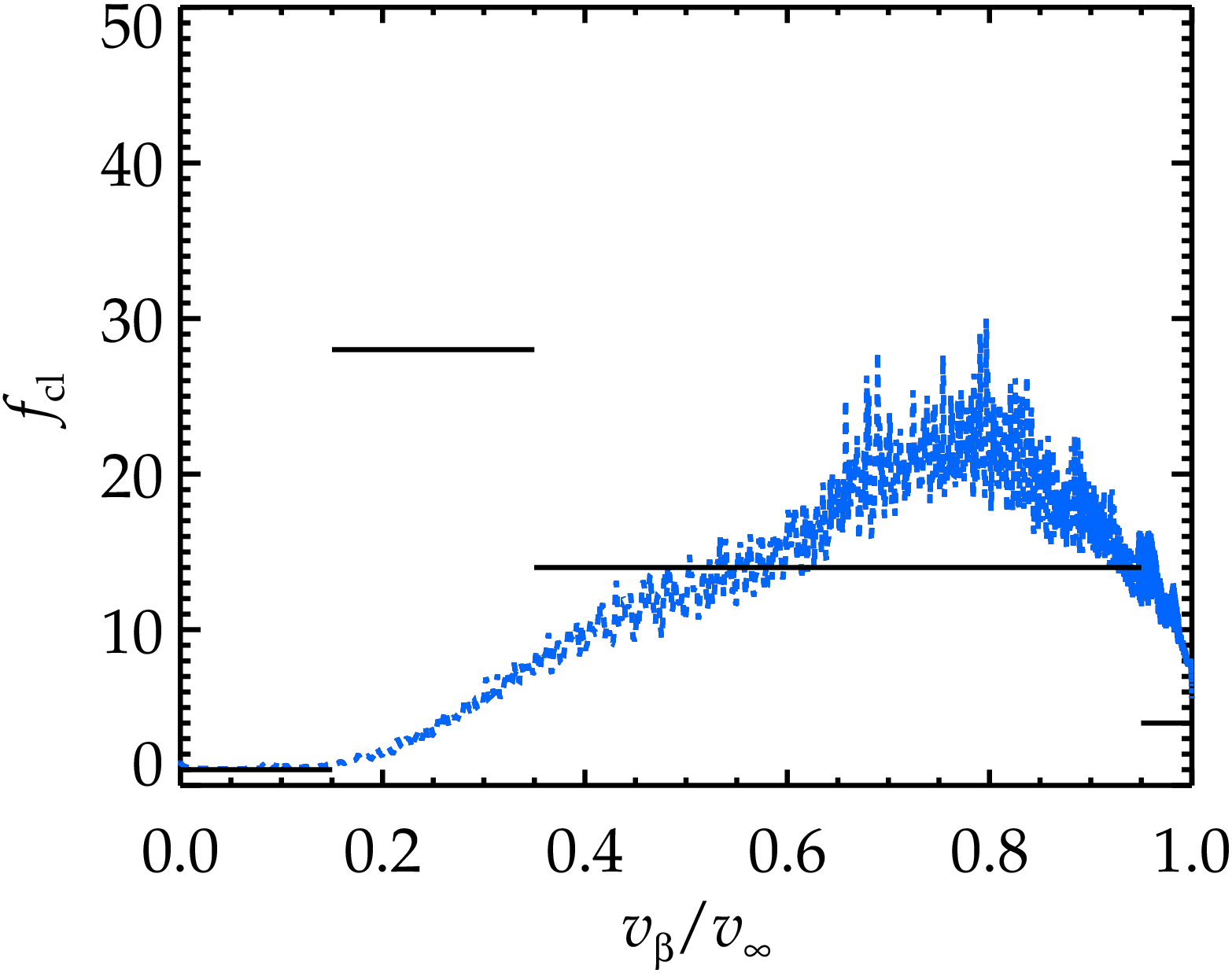}}
%\caption{This is the caption of the right-hand side figure. \label{fig_3}}
\end{minipage}
\caption{\textit{Left and middle panels:} Observed and modeled \pv~and
  \ha~line profiles in \lcep. Solid black lines are calculated from
  empirical, stochastic wind models, and blue dashed ones from RH
  simulations. The dashed-dotted line in the left panel illustrates
  corresponding results using the microclumping
  approximation. Observations (red dotted lines) are from Fullerton,
  et al.\ (2006) (\pv) and Markova et al.\ (2005) (\ha). 
  \textit{Right panel:} Corresponding clumping factors.}
\label{Fig:profs}
\end{figure}

We have carried out a combined \pv~and \ha~study, using both RH models
and stochastic ones. We find that synthetic spectra computed directly
from the RH models are unable to reproduce the diagnostic lines
(Fig.~\ref{Fig:profs}). Two main problems are identified: i) the
absorption toward the blue edge of \pv~is too deep, and ii) the core
emission of \ha~is much lower than observed. Note that changing the
mass-loss rate for which the RH model is calculated
(\mdot=1.5\,$\times$\,\mdotun) does not resolve these issues, or even
alleviate them; if for example a higher rate is adopted, the wings of
\ha~become much too strong.  Consequently, we apply our stochastic
models, aiming to empirically capture the essence of the structured
wind. By means of these models, we obtain consistent fits essentially
by increasing the clumping in the lower wind, but also by adopting
somewhat lower velocity spans of the clumps. The second point largely
resolves the issue of reproducing the observed \pv~lines, and was
extensively discussed in Sundqvist et al.\ (2010a).  Regarding the
first point, the observed \ha~absorption trough followed by the steep
incline to rather strong emission can only be reproduced by our models
if clumping is assumed to start at a velocity only marginally lower
than predicted by the RH models (see also Puls et al.\ 2006; Bouret et
al. 2008), but with a much steeper increase with velocity
(Fig.~\ref{Fig:profs}, right panel).  Regarding the outermost wind, RH
simulations by Runacres \& Owocki (2002), which extend to much larger
radii than those used here, indicate that the clumping factor there
settles at approximately four. \fcl\,$\approx$\,4 is consistent with
our derived mass-loss rate (see below) and the constraints from radio
emission derived by Puls et al.\ (2006), suggesting that the outermost
wind is better simulated than the inner by current RH models. However,
let us point out that the use of the so-called smooth source function
formalism in our RH models probably leads to overly damped
perturbations in the inner wind (e.g., Owocki \& Puls 1999), which
might at least partly explain the discrepancies between theoretical
clumping factors and those inferred from observations.

The basic expectations from the clump optical depth estimates in
Sect.~\ref{tau_cl} are confirmed by our detailed analysis. The
\pv~line profiles are clearly weaker than those calculated using
microclumping (Fig.~\ref{Fig:profs}, left panel), whereas \ha~is
affected only in the line core (not shown).  Finally, our derived
mass-loss rate for \lcep~is the same as the one used in our RH
simulations. It is approximately two times lower than predicted by the
line-driven wind theory, but a factor of five higher than the
corresponding rate derived assuming microclumping. This suggests that
only moderate reductions of current mass-loss predictions for OB-stars
might be necessary, and illustrates how a correct description of
clumping is pivotal for obtaining consistent, reliable estimates of
mass-loss rates.

%
% USE A SECTION WITHOUT NUMBER FOR THE ACKNOWLEDGEMENTS
%
\section*{Acknowledgements}
J.O.S and J.P acknowledge a travel grant from the DFG cluster of excellence.

% BEGIN THE REFERENCE LIST WITH \beginrefer
% USE \refer BEFORE THE REFERENCES AND BEGIN A NEW PARAGRAPH AFTER THE 
% REFERENCE !
% DO NOT FORGET TO END THE LIST WITH \endrefer
%
%\bibliographystyle{aa_mod}
%\bibliography{sundqvist_liege}

\footnotesize
%\begin{thebibliography}{23}
\beginrefer
%\expandafter\ifx\csname natexlab\endcsname\relax\def\natexlab#1{#1}\fi

\refer Bouret, J.-C., Lanz, T., \& Hillier, D.~J. 2005, \aap, 438, 301

\refer {Bouret}, J., {Lanz}, T., {Hillier}, D.~J., {et~al.} 2008, in Clumping
in Hot-Star Winds, ed. {W.-R.~Hamann, A.~Feldmeier, \& L.~M.~Oskinova}, 31

\refer {Castor}, J.~I., {Abbott}, D.~C., \& {Klein}, R.~I. 1975, \apj, 195, 157

\refer {Cohen}, D.~H., {Leutenegger}, M.~A., {Wollman}, E.~E., {et~al.} 2010, \mnras,
 405, 2391

\refer {Dessart}, L. \& {Owocki}, S.~P. 2002, \aap, 383, 1113

\refer {Feldmeier}, A., {Puls}, J., \& {Pauldrach}, A.~W.~A. 1997, \aap, 322, 878

\refer {Fullerton}, A.~W., {Massa}, D.~L., \& {Prinja}, R.~K. 2006, \apj, 637, 1025

\refer {Hamann}, W.-R., {Feldmeier}, A., \& {Oskinova}, L.~M., eds. 2008, {Clumping in
        hot-star winds}

\refer {Markova}, N., {Puls}, J., {Scuderi}, S., \& {Markov}, H. 2005, \aap, 440, 1133

%\refer {Oskinova}, L.~M., {Feldmeier}, A., \& {Hamann}, W.-R. 2006, \mnras, 372, 313

\refer {Oskinova}, L.~M., {Hamann}, W.-R., \& {Feldmeier}, A. 2007, \aap, 476, 1331

\refer {Owocki}, S.~P. 2008, in Clumping in Hot-Star Winds, ed. W.-R. {Hamann},
         A.~{Feldmeier}, \& L.~M. {Oskinova}, 121--

\refer {Owocki}, S.~P., {Gayley}, K.~G., \& {Shaviv}, N.~J. 2004, \apj, 616, 525

\refer {Owocki}, S.~P. \& {Puls}, J. 1999, \apj, 510, 355

\refer {Puls}, J., {Kudritzki}, R.-P., {Herrero}, A., {et~al.} 1996, \aap, 305, 171

\refer {Puls}, J., {Markova}, N., {Scuderi}, S., {et~al.} 2006, \aap, 454, 625

\refer {Puls}, J., {Urbaneja}, M.~A., {Venero}, R., {et~al.} 2005, \aap, 435, 669

\refer {Puls}, J., {Vink}, J.~S., \& {Najarro}, F. 2008, \aapr, 16, 209

\refer {Runacres}, M.~C. \& {Owocki}, S.~P. 2002, \aap, 381, 1015

\refer {Sundqvist}, J.~O., {Puls}, J., \& {Feldmeier}, A. 2010a, \aap, 510, A11

\refer {Sundqvist}, J.~O., {Puls}, J., {Feldmeier}, A., {et~al.} 2010b, submitted to~\aap

\refer {Vink}, J.~S., {de Koter}, A., \& {Lamers}, H.~J.~G.~L.~M. 2000, \aap, 362, 295

\endrefer           

\end{document}